\documentclass[]{aiaa-tc}

\usepackage{amsmath}
\usepackage{amssymb}
\usepackage{doi}
\usepackage{lettrine}
\usepackage{subfig}
\usepackage{url}
\usepackage{wrapfig}
\usepackage{color}
\definecolor{blue}{rgb}{0.098,0.357,0.675}
\definecolor{green}{rgb}{0.5,0.75,0.0}

\usepackage[]{hyperref}
\hypersetup{
    pdftex, 
    pdftitle={AIAA paper}, 
    pdfauthor={J. Brent Parham, Lorena A. Barba}, 
    pdfpagemode={UseOutlines},
    bookmarks, bookmarksopen,bookmarksnumbered={True},
    colorlinks, linkcolor={blue},citecolor={blue},urlcolor={blue}}

\newcommand{\vecb}[1]{\mathbf{#1}}

\newcommand\ignore[1]{} 
\newcommand{\comment}[1]{\textcolor{red}{ #1 }}
\ignore{
\renewcommand{\comment}[1]{}
}

\title{Finding the Force---Consistent Particle Seeding for Satellite Aerodynamics}
 \author{
  J. Brent Parham\thanks{Lincoln Scholar at Boston University, Department of Mechanical Engineering; AIAA Member.}\\
  {\normalsize\itshape
   MIT Lincoln Laboratory, Lexington, MA 02421, U.S.A.}\\
  \and
  L.~A. Barba\thanks{New address: Mechanical and Aerospace Engineering, George Washington University; AIAA Member.}\\
  {\normalsize\itshape
 Mechanical Engineering, Boston University, Boston, MA 02215, U.S.A.}
 }
 
\begin{document}

\maketitle

\begin{abstract} 
\let\thefootnote\relax\footnote{\\ \bf\centering Distribution Statement A. Approved for public release; distribution is unlimited.}

When calculating satellite trajectories in low-earth orbit, engineers need to adequately estimate aerodynamic forces. But to this day, obtaining the drag acting on the complicated shapes of modern spacecraft suffers from many sources of error. While part of the problem is the uncertain density in the upper atmosphere, this works focuses on improving the modeling of interacting rarified gases and satellite surfaces. The only numerical approach that currently captures effects in this flow regime---like self-shadowing and multiple molecular reflections---is known as test-particle Monte Carlo. This method executes a ray-tracing algorithm to follow particles that pass through a control volume containing the spacecraft and accumulates the momentum transfer to the body surfaces. Statistical fluctuations inherent in the approach demand particle numbers in the order of millions, often making this scheme too costly to be practical. This work presents a parallel test-particle Monte Carlo method that takes advantage of both GPUs and multi-core CPUs. The speed at which this model can run with millions of particles allowed exploring a regime where a flaw in the modelÕs initial particle seeding was revealed. Our new model introduces an analytical fix based on seeding the calculation with an initial distribution of particles at the boundary of a spherical control volume and computing the integral for the correct number flux. This work includes verification of the proposed model using analytical solutions for several simple geometries and demonstrates uses for studying aero-stabilization of the Phobos-Grunt Martian probe and pose-estimation for the ICESat mission.

\end{abstract}

\section*{Nomenclature}

\begin{tabbing}
  XXX \= \kill
  $k_B$ \>= Boltzmann constant, J/K\\  
  $T$ \>= Temperature, K\\
  $m$ \>= Molecular mass of gas particles, kg\\
  $A_\text{ref}$ \>= Reference Area, m$^2$\\
  $n$ \>= Number density, \#/m$^3$\\
  $\rho$ \>= mass density ($nm$), kg/m$^3$\\
  $C_D$ \>= Drag coefficient, (Drag Force)/$\left(\frac{1}{2}\rho V^2A_{ref}\right)$\\
  $\Gamma$ \>= Number flux through surface \\
  $c_{mp}$ \>= Most probable thermal speed, $\sqrt{2{k_B}T_\infty/m}$\\
  $S$ \>= Speed ratio, $V/c_{mp}$\\
  \\
  \emph{Subscripts}\\
  $\infty$ \>= Free stream equilibrium conditions\\
  $W$ \>= Properties evaluated at the satellite surface\\
\end{tabbing}
 
\setcounter{footnote}{0}

\section{Introduction}

\lettrine[nindent=0pt]{N}{early five years ago}, the active communications satellite Iridium 33 crashed into Cosmos 2251---a defunct Russian satellite---scattering thousands of satellite fragments into low-earth orbit.\footnote{A software reconstruction of the collision is available on video at \url{http://youtu.be/_o7EKlqCE20}, courtesy of AGI.} This major space collision, among other similar events, alerted satellite operators to the dangers of space-debris proliferation and motivated world-wide response \cite{un2007}.  With a sharp increase in new objects to track in space, accurately estimating the orbits and trajectories of satellites while reducing uncertainties is now a critical aspect of preventing future collisions and reducing the cost of maneuvers to evade debris.   

Just a year before the Iridium 33 crash, Vallado \cite{vallado2008} had pointed out that a lack of research in the area of physically consistent drag estimation was an obstacle for predicting satellite orbits accurately.  The situation remains the same today. Aerodynamic forces induce the most uncertain accelerations encountered by a satellite in Low-Earth Orbit (LEO), but they remain elusive to calculate.  The drag coefficient itself is a geometry-dependent  factor multiplying the dynamic pressure to give the drag force, as shown in equation \eqref{eq:drag}, with atmospheric neutral density $\rho$, and orbital velocity $V$.
\begin{align}
F_{\text{Drag}}={C_D A_\text{ref}}\frac{1}{2}\rho V^2 \label{eq:drag}
\end{align}

\noindent The main uncertainties in this equation arise with specifying the neutral density and calculating the drag area ($ C_D A_\text{ref}$) for a given satellite, which can have an arbitrarily complicated geometry.  While recent work has improved the accuracy of atmospheric neutral density models \cite{sutton2009,marcos2010} by inferring densities from satellite tracking data---e.g., the HASDM model \cite{Storz2005}---, calculating satellite coefficients numerically in the context of orbit determination needs to be revisited and standardized.  Uncertainty reduction in \emph{all} terms of the drag force is the only way to increase the precision of its estimate.

The earliest attempts to model satellite aerodynamics were analytical \cite{sentman1961,chambre1961}, with aerodynamic forces directly integrated from momentum flux over a surface. But closed-form solutions were only obtained for simple geometries: flat plate, cone, cylinder and sphere.  These solutions omitted effects such as shadowing and multiple reflections, which may change aerodynamic properties.  As computational power grew through the years, the Test-Particle Monte Carlo (TMPC) method \cite{bogacheva1969monte,bird1994,klinkrad1995} appeared as an alternative to analytical models. Test-particle methods trace simulated particles through a control volume that contains the spacecraft.  Each particle can then interact with the surface to transfer momentum and energy through a surface interaction model (e.g., diffuse, specular, etc.). The procedure reduces to a Monte Carlo initialization and a geometric ray-tracing algorithm.  TPMC methods remain the only way to adequately account for shadowing and multiple reflections with arbitrary surface interaction. But, like all Monte Carlo methods, TPMC needs a large number of particles to reduce the statistical scatter of the solutions. The computational cost therefore severely limits its application to orbital analysis. 

A TPMC simulation is set up by placing particles randomly with a uniform distribution on the faces of a rectangular control volume.  This initialization, however, is inconsistent with kinetic theory for arbitrary free-stream incidence.  Here, we construct a new method to initialize a TPMC simulation that solves this inconsistency.  By choosing a spherical control volume, we reveal an exact solution for initial-position likelihood of incident particles and the total number flux into the volume.  We developed a code implementing TPMC, then extended it to include the new method and to run in parallel using modern computational hardware---both multi-core CPUs and many-core GPUs.  In this paper, we describe the mathematical derivation for the physically consistent test-particle seeding along with the validation of the method and verification of the code using analytical tests. We also demonstrate the application of the new method to orbital analysis with actual satellite data.

\section{Methods}


\subsection{Test-particle Monte Carlo (TPMC) method}

A rarefied-gas flow is well modeled by a drifting Maxwellian gas, i.e., a gas described by the following velocity distribution,
\begin{align}
f(v_i) &= \sqrt{\frac{m}{2\pi k_B T}}\exp\left[-\frac{m(v_i-u_i)^2}{2k_BT}\right],\label{eq:maxwellian}
\end{align}

\noindent where the bulk average velocity components, ${u}_i$, are superimposed on the individual distribution components, $v_i$ and an isotropic particle density distribution in space is assumed, $f(x_i) = \text{constant}$.

In free molecular flow---where the mean free path of molecules is large in relation to the volume considered---the gas particles do not interact with each other in the free stream nor with the particle distribution created by surface reflections.  The TPMC method takes advantage of this limit by ignoring particle-particle interactions and treating particles independently. It then obtains  the average momentum transfer to a body from the sum of discrete particle reflections.  ``Particles'' here are meant to represent a large number of gas molecules, for which a weighting factor is introduced. Using test-particle methods requires accurate sampling of the free-stream flux into the specified control volume and a model of particle interactions with body surfaces.

Fritsche and Klinkrad \cite{Fritsche:2004uq} explain a flux-sampling scheme of the TPMC method implemented in the  software that is used by the European Space Agency, called ANGARA.  The sampling consists of placing particles with a random uniform distribution on the six faces of a rectangular control volume. The number of particles to cross each face derives from the analytical molecular flux across a flat plate with given normal velocity component of the free stream and the total number of particles in the TPMC simulation (a parameter dictating accuracy).  
Once the particles are initialized on the control surface, they are followed via a ray-tracing algorithm until they intersect with the surface of the spacecraft. Satellite surfaces are represented by a triangulation, allowing the use of well-known ray-tracing methods.  To account for shadowing, the closest intersection point to the initial position becomes the first collision with the satellite and the particle is moved to this point.  At the new location,  a surface-interaction model determines the new velocity for the particle.

The surface-interaction model most used in satellite aerodynamics is diffuse reflection,
 which is both simple and has been proven to adequately model satellite surface interactions in LEO \cite{moe2011}.  In this model, the incident particles settle in the surface material and are reemitted with a velocity distribution determined by the satellite surface temperature.  In TPMC, a random departure direction from the surface is specified and the reemission speed is chosen from the Maxwellian speed distribution (the product of the three components in equation \eqref{eq:maxwellian} integrated over all angles with spherical coordinates in velocity space).   
The difference between the original and final velocities are then scaled by the particle's effective mass and added to the total momentum exchange with the spacecraft.  To calculate the particle's effective mass, the total number of incident molecules, derived from the molecular flux, is divided by the number of simulated particles (specified by the user) and multiplied by the molecular mass.  The final resultant aerodynamic  force is then the total momentum exchange that includes contributions from all particles.  

\subsection{Analytical solutions for simple geometries}

The following three analytical solutions provide a way to scrutinize the current model.  The solutions assume a diffuse reemission and the resultant force derives from a surface integral of the momentum flux on the satellite surface.  The influx of momentum is integrated using the free-stream distribution and the outflux is calculated with the reemission distribution described by the satellite temperature. To simplify, we define the ratio of the free-stream speed magnitude, $U_\infty$, to the most-probable thermal speed, $c_{mp}=\sqrt{2{k_B}T_\infty/m}$, as the ``speed ratio,"   $S=U_\infty/c_{mp}$.

\subsubsection{Flat plate at angle of incidence}

This solution apears in Sentman's work.\cite{sentman1961} The first three terms represent the influx of momentum integrated over a planar surface and the last term is the reemission contribution to the total momentum flux. The drag coefficient is then given by
\begin{align}
C_D=\frac{2}{\sqrt{\pi}S} \left( \exp[-(S\cos\alpha)^2]+\sqrt{\pi}S\cos\alpha \left(1+\frac{1}{2S^2}\right)\text{erf}[S\cos\alpha]+\frac{\pi S}{S_W}\cos^2\alpha \right), \label{eq:plate}
\end{align}

\noindent  where $S_W$ is the speed ratio using the wall temperature of the body in $c_{mp}$ and $\alpha$ is the angle of attack.

\subsubsection{Convex spherical segment}

As given by Bird \cite{bird1994}, this result is recreated by integrating equation \eqref{eq:plate} over a spherical surface with the appropriate change in angle of attack and choice of differential area element. The drag coefficient is
\begin{align}
C_D=\frac{\exp[-S^2]}{\sqrt{\pi}S^3}(1+2S^2) + \frac{4S^4+4S^2-1}{2S^4}\text{erf}[S] + \frac{2\sqrt{\pi}}{3S_W}. \label{eq:sphere}
\end{align}

\subsubsection{Concave spherical segment in hyperthermal flow}

Pratts's thesis on concave bodies in hyperthermal flow \cite{pratt1963} generalizes the drag coefficient of a hemispherical body with full thermal accommodation to the following expression (in his notation):
\begin{align}
C_D=2+\epsilon_D(\theta)\frac{\sqrt{\pi}}{S}\sqrt{\frac{T_W}{T_\infty}}. \label{eq:hypersphere}
\end{align} 

\noindent Here, $\epsilon_D(\theta=90^\circ)=1.05349$ is a numerical solution to the integral that Pratt uses for the modified ``reemission drag'' of a concave hemispherical surface.  Though this analytical approximation is limited, it provides a good comparison to make sure concave effects are exhibiting qualitatively correct behavior.

\subsection{Satellite-tracking derived coefficients}

The accelerations caused by drag forces can manifest themselves in tracking data of satellites.  By tracking satellites for a length of time, the orbital state and ballistic coefficient are derived with an orbit-determination routine, most commonly a least-squares optimization.  The North American Aerospace Defense Command (NORAD) releases the resulting states publicly through the web service Space-Track.\footnote{\url{https://www.space-track.org}}  Vallado standardized the details of the Simplified General Perturbations (SGP4) dynamical model used to compute these two-line element sets (TLEs) with his interpretation of the original specifications \cite{Vallado:2006uq}. We use his interpretation for the analysis of TLEs.  

For an LEO object, the TLE contains information about the drag force during the period that the element set was created. This ballistic coefficient $B^*$ of the TLE maps directly to the the drag area ($R_\oplus=6378.1$ km and $\rho_0=2.461 \times 10^{-5}$ kg/m$^2$):
\begin{align}
C_DA_{ref} = \frac{2B^*M_{\text{satellite}}}{R_\oplus\rho_0} \label{eq:tlebstar}
\end{align}

The International Laser Ranging Service (ILRS)\footnote{\url{http://ilrs.gsfc.nasa.gov/data_and_products}} also releases tracking data for participating satellites.  An orbit determination routine---along with an initial orbit estimate from a TLE---creates a precise orbital state with this range data. The precision estimate serves as a check to the TLE data for the ICESat mission and is processed with an orbit determination routine based on the work of Montenbruck and Gill \cite{montenbruck2000}. 

\section{Results}

\subsection{Physically consistent particle seeding for TPMC}

\begin{figure}[htbp]
\begin{center}
\includegraphics[width=3.5in]{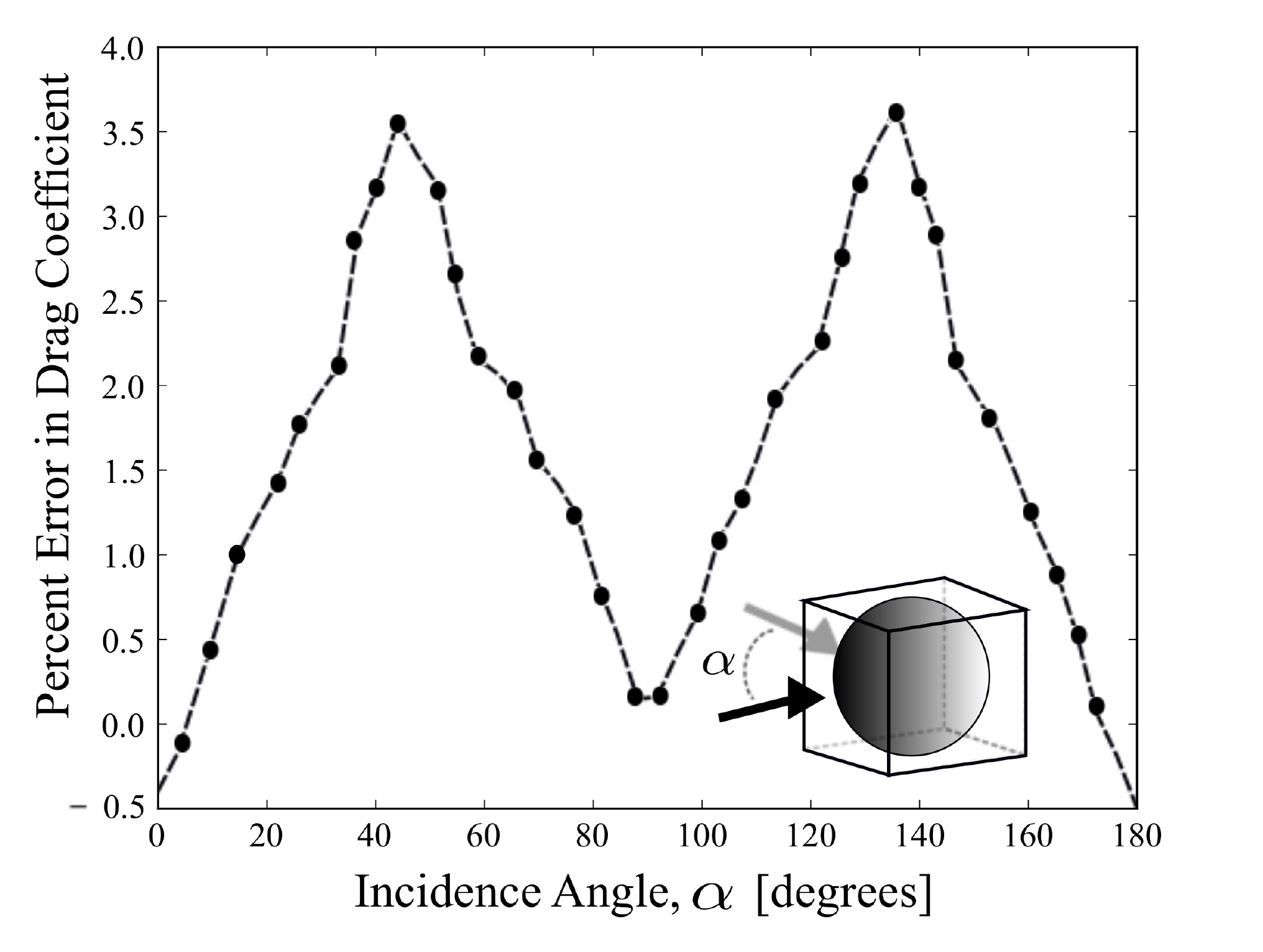}
\caption{{  Percent error of a 10-million-particle TPMC calculation with respect to exact sphere drag.  Both models assume fully diffuse reflection.  The TPMC  result is not a constant function of incidence angle to the rectangular control volume. }}
\label{fig:sphere_err}
\end{center}
\end{figure}

TPMC simulations begin by placing a uniform random distribution of particles on the faces of a rectangular domain.  The increasing demands---{i.e.}, accuracy for orbital analysis---expose the biases introduced by seeding on a rectangular domain: the method constructs spurious high-density regions at the edges of the domain faces.  As a result, the drag coefficient (in this test, of the sphere) shows an artificial increase as the free stream is swept around the rectangular control volume (see Figure \ref{fig:sphere_err}).  

To avoid these density artifacts, we modified the TPMC method by developing a particle-seeding scheme on a spherical control volume.  We obtained closed-form solutions for this control volume by solving the associated probability integrals for a drifting Maxwell gas. The solution gives the correct number flux and particle distribution on the spherical shell, which in turn gives the correct particle weights, sampled positions and sampled velocities.

\subsubsection{Position sampling}

The number flux and particle distribution are set up as an integral of the velocity distribution along the spherical surface.  We start by defining a flux, $\Gamma$, of particles \emph{into} a surface, with a given velocity distribution $f(\vecb{v})$, and write the following expression (where $n$ is the number density):

\begin{align}
\Gamma = n\int_\text{Velocity} \left[\int_\text{Area} \vecb{v}f(\vecb{v})\cdot d\vecb{a}\right]d\vecb{v}. \label{eq:flux}
\end{align}

\begin{wrapfigure}[17]{L}{0.4\linewidth}
\includegraphics[]{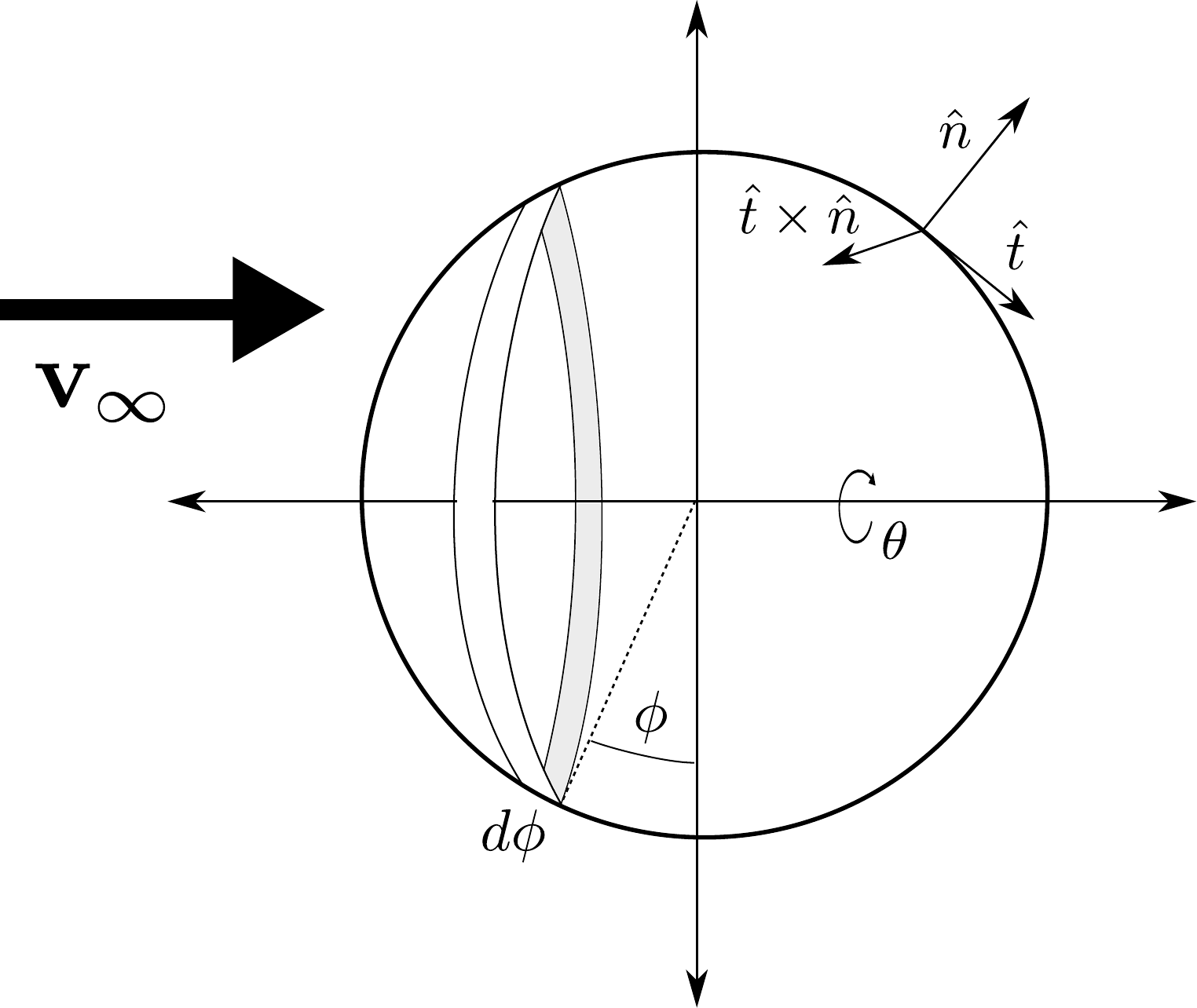}
\caption{{ Spherical geometry used to solve molecular influx to the control volume.}}
\label{fig:sph_geo}
\end{wrapfigure}

We consider a set of orthogonal coordinates that are aligned with the normal and tangent vectors, as shown in Figure \ref{fig:sph_geo} (from here onwards $\hat{t}_1=\hat{t}$ and $\hat{t}_2=\hat{t}\times\hat{n}$).  Aligning the inward velocity with the spherical normal, we rewrite the integral as shown in equation \eqref{eq:integration}.  The bulk velocity components $u,v,$ and $w$ are expressed in the local coordinate system.

\begin{align}
\Gamma&=\frac{n}{(c_{mp}^2\pi)^{3/2}}\int_\text{Area}\left[\int_{0}^{\infty}\int_{-\infty}^{\infty}\int_{-\infty}^{\infty} {v_n} \exp \left(-\frac{({v_n}-u)^2+({v_{{t_2}}}-v)^2+({v_{t_1}}-w)^2}{c_{mp}^2}\right) dv_{t_1} dv_{{t_2}}dv_n \right]da \notag\\
&=\frac{n}{(c_{mp}^2\pi)^{3/2}}\int_{\phi=-\pi/2}^{\pi/2}\int_{\theta=0}^{2\pi} \frac{1}{2} c_{mp}^2 \pi \left( c_{mp}^2 e^{-\frac{u^2}{ c_{mp}^2}}+{\sqrt{\pi } u c_{mp}}\left[1 + \text{erf}\left(\frac{u}{c_{mp}}\right)\right]\right)r^2 \cos\phi d\theta d\phi \label{eq:integration}\\
\notag
\end{align}

The free stream is aligned with the $\{\phi,\theta\}=\{\pi/2,0\}$ direction so that only one component remains in specified spherical coordinates ($u=v \sin\phi$), and we can integrate to find the result:

\begin{align}
\Gamma=\frac{nr^2}{(c_{mp}^2\pi)^{3/2}}\left[\pi ^2 c_{mp}^4 e^{-\frac{v^2}{c_{mp}^2}}+\frac{\pi ^{5/2} c_{mp}^3 \left(c_{mp}^2+2 v^2\right) \text{erf}\left(\frac{v}{c_{mp}}\right)}{2 v}\right].
\end{align}

Simplifying further, by assuming that both $v$ and $c_{mp}$ are positive and defining the speed ratio as $S= v/c_{mp}$, we obtain

\begin{align}
\Gamma&= \left[\sqrt{\pi} c_{mp}e^{-\frac{v^2}{c_{mp}^2}}+\left(\frac{\pi c_{mp}^2}{2v}+\pi v\right)\text{erf}\left(\frac{v}{c_{mp}}\right)\right]nr^2 \notag\\
&=\left[\sqrt{\pi}e^{-S^2}+\left(\frac{\pi}{2S}+\pi S\right)\text{erf}\left(S\right)\right]c_{mp}nr^2.
\end{align}

This number flux is an exact expression for the flux into a sphere for all Maxwellian gases in equilibrium.  With this result, a likelihood $g(\phi|S)$ of incidence angle $\phi\in[-\pi/2,\pi/2]$ into a sphere can be constructed and characterized only by the speed ratio, as follows:

\begin{align}
g(\phi|S) = \
\frac{ \cos\phi\left[e^{-S^2 \sin^2\phi} +  S\sqrt{\pi}\sin\phi\left( 1 + \text{erf}(S\sin\phi)\right) \right]}{ e^{-S^2} + \frac{\sqrt{\pi}(1 + 2S^2)\text{erf}(S)}{2S}}.
\end{align}

\begin{figure}
\centering
	\subfloat[Likelihood for distribution of particles based on angle of incidence and various speed ratios.]{\includegraphics[width=0.6\linewidth]{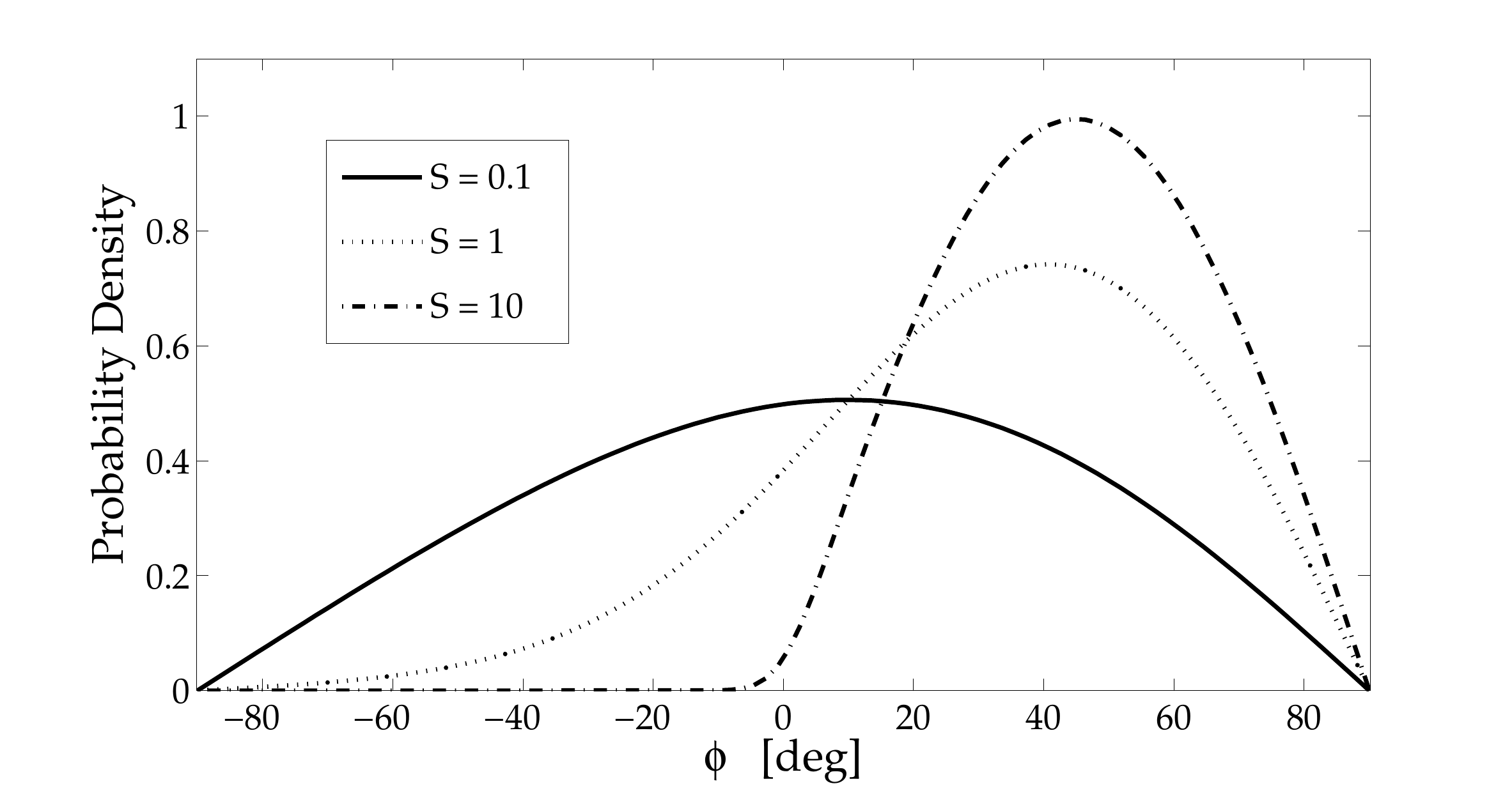}\label{fig:speed}}\\
	\subfloat[Initial position sampling of 10,000 particles for various speed ratios]{\includegraphics[width=0.7\linewidth]{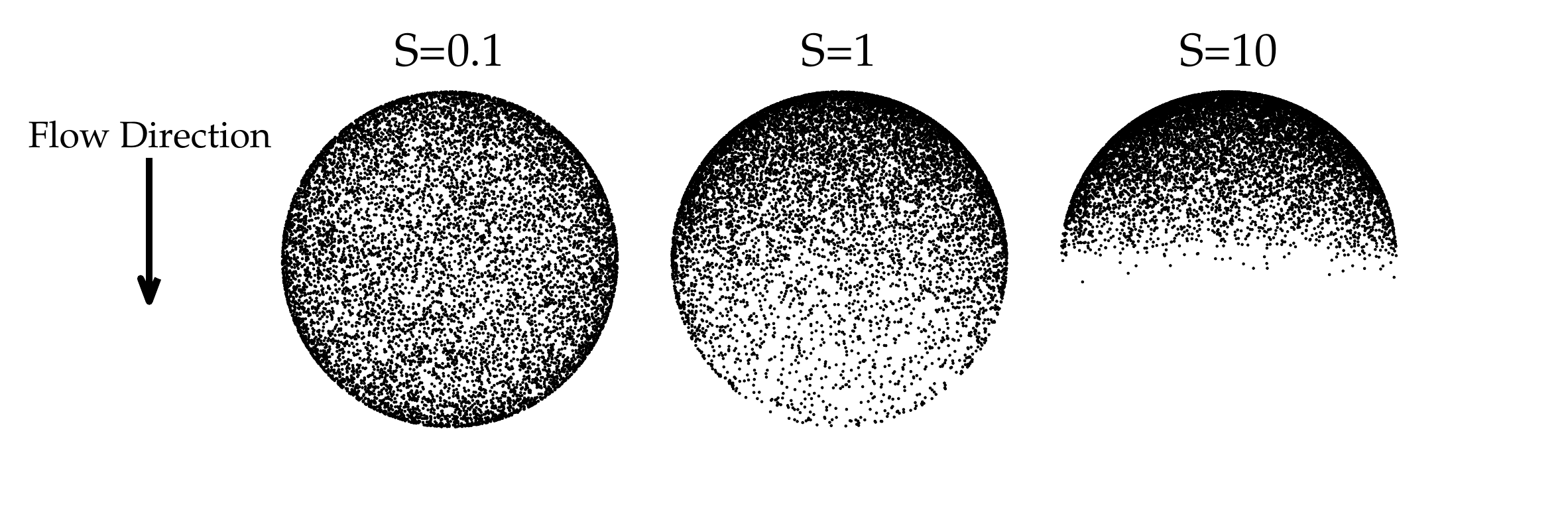}\label{fig:speed_sample}}
 \caption{Physically consistent particle sampling, applied to a spherical volume.}\label{fig:sample}
\end{figure}

The evaluation of this distribution for several speed ratios is plotted in Figure \ref{fig:speed}.  For low speed ratio the distribution approaches the limit
\begin{align}
\lim_{S\rightarrow0} g(\phi|S)=\frac{1}{2}\cos\phi. \label{eq:lowS}
\end{align}

When the density in equation \eqref{eq:lowS} is combined with a uniform distribution sampling $\theta\in[0,2\pi]$, we obtain a uniform sampling on a spherical surface.  
This limit exhibits the spatially isotropic thermal influxes from a resting gas that is in equilibrium.  
With large speed ratio, the distribution approaches 
\begin{align}
\lim_{S\rightarrow\infty} g(\phi|S_\infty)= 
\begin{cases}
\sin(2\phi) & \phi \geq 0\\
 0 & \phi<0
\end{cases}
\end{align}

This is the hyper-thermal limit where particles are distributed uniformly on a surface with a normal direction aligned with the free stream. 
The free stream velocity---not the thermal fluctuations---dictate the influx (Figure \ref{fig:speed_sample}).  
Once the two angles ($\phi$ and $\theta$) exist, we have succeeded and can write the position of the particle in Cartesian space.

\subsubsection{Velocity sampling}

With a position sampled, we choose the velocity using the canonical velocity component distributions for flux past a flat plate, {i.e.}, a surface element with the normal aligned with the position vector.  The distributions in velocity space transform into the local coordinate frame with respect to the sphere's normal vector, and are represented in Equations \eqref{eq:vel_dist1}, \eqref{eq:vel_dist2} and \eqref{eq:vel_dist3}.

\begin{subequations}
\begin{align}
p(v_n) &= \frac{2v_i\exp[-\left(\frac{{v}_n}{c_{mp}}-{S}_n\right)^2]}{\exp[-{S}_n^2]+\sqrt{\pi}{S}_n(1+\text{erf}[{S}_n])} \label{eq:vel_dist1}\\
p(v_{t_1}) &= \frac{1}{\sqrt{\pi}}\exp\left[-\left(\frac{{v}_{t_1}}{c_{mp}}-{S}_{t_1}\right)^2\right] \label{eq:vel_dist2}\\
p(v_{{t_2}}) &= \frac{1}{\sqrt{\pi}}\exp\left[-\left(\frac{{v}_{{t_2}}}{c_{mp}}-{S}_{{t_2}}\right)^2\right] \label{eq:vel_dist3}
\end{align}
\end{subequations}

\noindent The free stream velocity component-wise speed-ratio in the local coordinate system can be written as, $S_{t_1} = |\vecb{S}|\cos\phi$, and $S_n = |\vecb{S}|\sin\phi$ (inwardly pointing here) leaving the third component, $S_{t_2}$, always equal to zero. 

\subsubsection{Rotation to body coordinates}

The position vectors are rotated from the free-stream centered frame, with axis $\tilde{z}$, back into the satellite reference frame, with axis $\hat{z}$, with a rotation about the vector axis $\hat{u}=\tilde{z}\times\hat{z}$ by the angle $\beta=\cos^{-1}(\tilde{z}\cdot\hat{z})$. This rotation is implemented using Rodrigues' rotation formula, given as:

\begin{align}
\vecb{x}^\prime = \vecb{x}\cos\beta+(\hat{u}\times\vecb{x})\sin\beta + \hat{u}(\hat{u}\cdot\vecb{x})(1-\cos\beta)
\end{align}

\noindent The velocity components transform back to the satellite reference frame via multiplication by the transpose of the local coordinate matrix, consisting of the vectors $\hat{n},\hat{t}_1$ and $\hat{t}_2$.

\begin{figure}
\centering
	\subfloat[NVIDIA K20 GPU compared to Intel Core i7 CPU]{\includegraphics[width=2.8in]{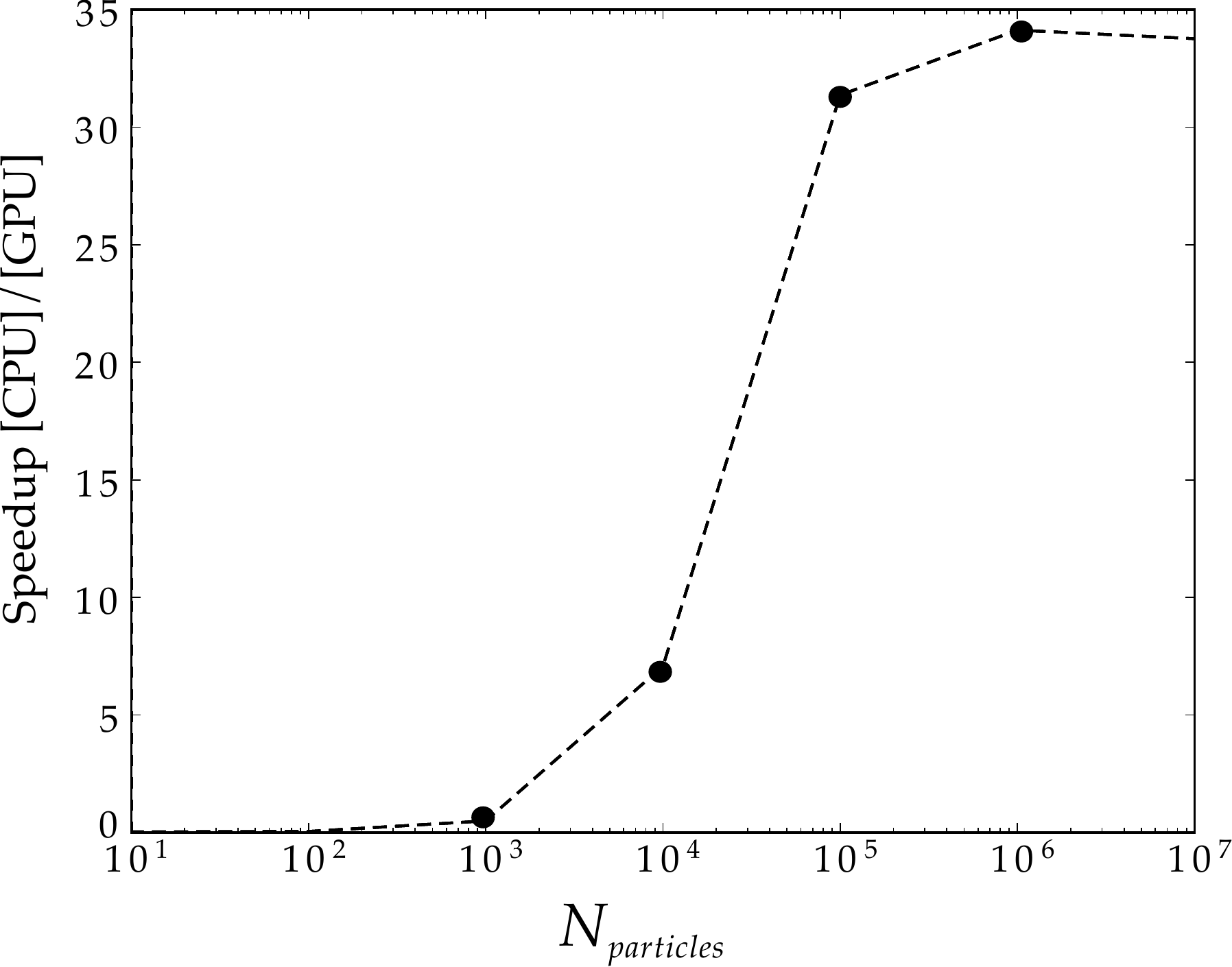}\label{fig:speedup}}\hspace{.1in}
	\subfloat[Decreasing statistical scatter with more particles]{\includegraphics[width=3.2in]{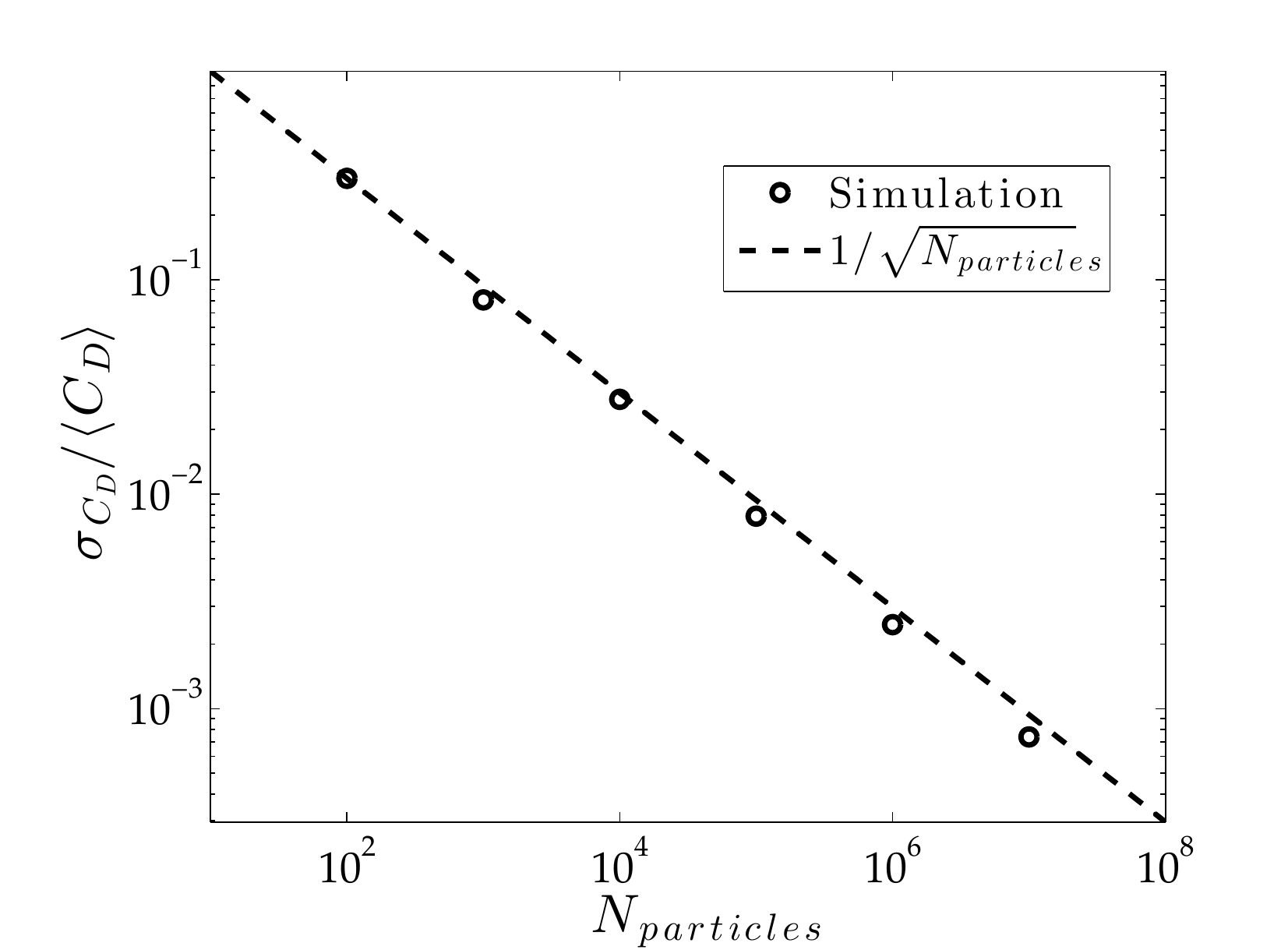}\label{fig:scatter}}
 \caption{Simulation workload and convergence characterizations}
\end{figure}

\subsection{Software implementation and characterization}

We implemented the revised TPMC ray-tracing method in CUDA---parallelized over particles---and tested on a Kepler-architecture NVIDIA Tesla K20 GPU. We also wrote an OpenMP version to run efficiently on an Intel Core i7 CPU. Chip-to-chip comparison shows up to 35-times speedup on the GPU, with appropriate problem size (Figure \ref{fig:speedup}). For all calculations in the following results, we ran the CUDA version with a particle count of 10 million (providing a statistical scatter of approximately 0.07\% for the comparison with the flat-plate analytical solution, as shown in Figure \ref{fig:scatter}).

\subsection{Verification with analytical solutions}

The comparison with analytical solutions allows a strict verification of the new boundary conditions and code implementations against free-molecular theory.  For the cases shown,  we set the free-stream temperature to $T_\infty=922$ K and the body-surface temperature to $T_W=300$ K. These conditions are typical of those experienced at an altitude of about 400 km.  
The triangular surface mesh of the simple geometric shapes were created using \verb=Gmsh=\cite{gmsh2009}.  

\begin{figure}[htbp]
\begin{center}
\includegraphics[width=3.5in]{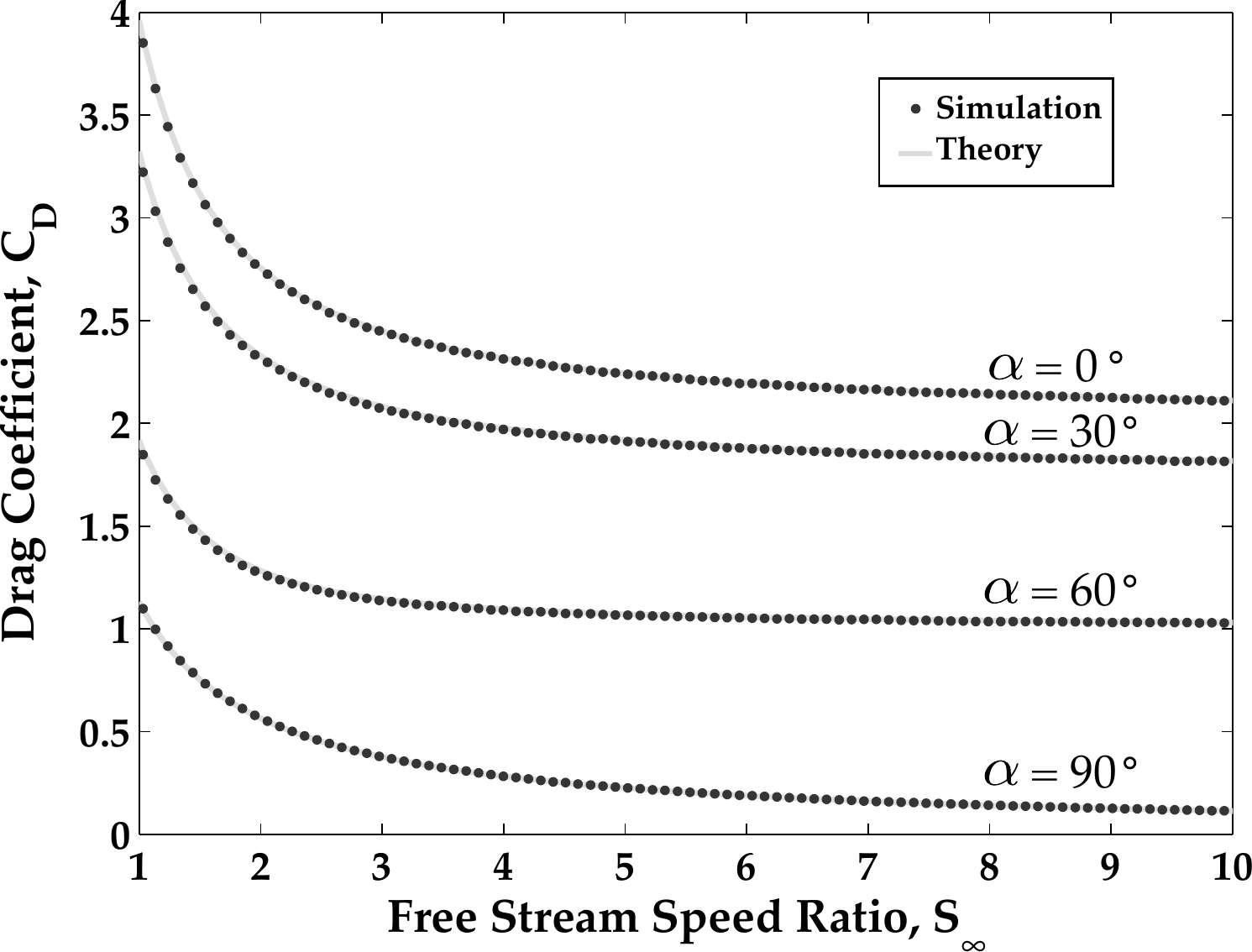}
\caption{The analytical solution of a double-sided flat plate at angle of attack compares favorably to the output of the TPMC code developed.  In this figure, free stream speed ratio and angle of attack are varied to show the dependance of the drag coefficient on those variables. }
\label{fig:flat_plate}
\end{center}
\end{figure}

Figure \ref{fig:flat_plate} shows that when the code is compared to the analytical solution of a double-sided flat plate (i.e., a plate that can absorb free-stream molecules from the front and back), our model exactly captures the variation with the speed ratio from near-thermal flows speeds to the extremely fast flows encountered by spacecraft.  As angle of attack changes from normal to the plate ($\alpha=0^\circ$) to parallel, $C_D$ decreases as expected.  The thermal fluctuations near the conditions at altitude ($S_\infty=7$) create a non-zero drag even on surfaces that are parallel to the flow.  This effect often causes underestimates in drag forces when the hyper-thermal approximation is used for analysis.  The full TPMC method accurately reproduces this physical phenomenon and can readily be extended to more complicated bodies.

\begin{figure}[htbp]
\begin{center}
\includegraphics[width=3.5in]{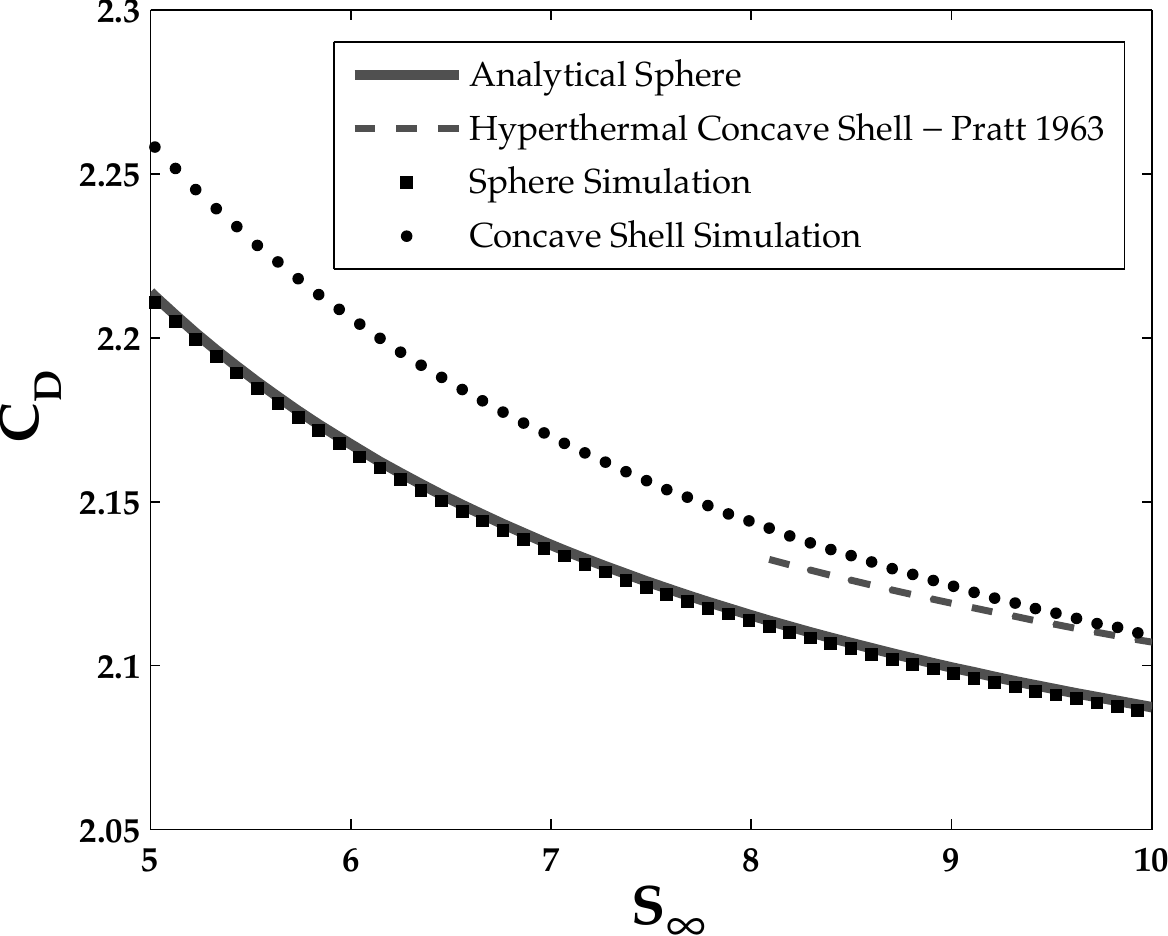}
\caption{The simulation for a sphere with varying speed ratio compared to its analytical solution.  The solution of a concave spherical shell is also shown, and it asymptotes to the hyper-thermal approximation given by Pratt.}
\label{fig:sphere_shell-CD}
\end{center}
\end{figure}

To verify the method against a slightly more complex body, we turn to the analytical solution for a sphere, as seen in Figure \ref{fig:sphere_shell-CD}.  Again, the physical variation with the speed ratio is captured by the TPMC method and we can confidently say that the code reproduces forces on convex surfaces. As a check to see if the code can accurately model the effects of multiple reflections, we compare Pratt's hyper-thermal model to the TPMC calculation of a concave shell over the whole domain of speed ratio (also in Figure \ref{fig:sphere_shell-CD}).  The TPMC solution asymptotically approaches the hyper-thermal solution as $S_\infty$ increases.  Although there is no direct solution of the reflection integrals for the whole domain, we can reasonably draw a conclusion that this verifies the reflection algorithm implemented in the method. 

\newpage

\subsection{Demonstration with satellite-tracking data}

\begin{wrapfigure}[25]{R}{0.55\linewidth}
\includegraphics{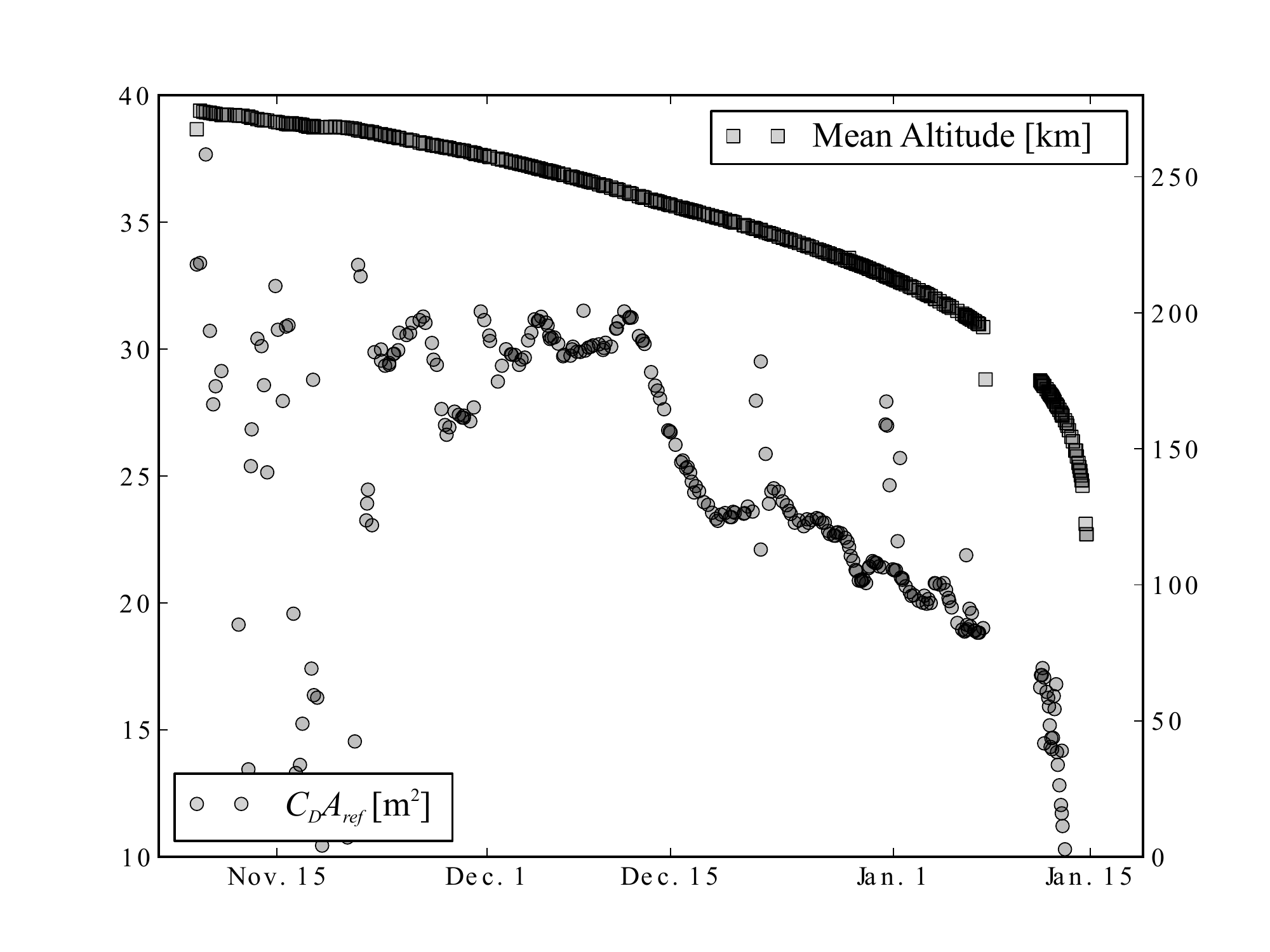}
\caption{The Phobos-Grunt probe aero-stabilized by the middle of December 2011, and Two Line Element (TLE) set derived force coefficients allow for comparison of model estimates to actual dynamical characteristics during this stabilization.}
\label{fig:PG}
\end{wrapfigure}

\subsubsection{Phobos-Grunt aero-stabilization}

Phobos-Grunt, a multinational scientific mission that failed to escape its initial parking orbit, reentered in early January 2012.  This failure provided a unique opportunity to observe a large spacecraft's uncontrolled interactions with the near-Earth environment.  From its launch in November 2011 to its reentry, the craft was observed by many amateur astronomers\footnote{\url{http://legault.perso.sfr.fr/phobos-grunt.html}} and appeared to aero-stabilize in mid-December 2011.  This behavior also appears in TLE-derived force coefficients (Figure \ref{fig:PG}) and can be compared to an estimated model of the craft's geometry.  Using a model based on Phobos-Grunt, we can simulate the torques encountered at altitude---with conditions specified in Table \ref{tab:PGcond}.  

All moments are shifted to a ``quarter-chord'' position that is measured from the fuel tank end. The total length of the craft, $c\approx5.5$ m, is taken to be the chord length.  There is good reason to believe that the center of mass is close to this point, since the tanks were still filled with a large amount of fuel meant for the orbital transfer to mars.

\begin{table}[htdp]
\caption{Flight conditions for a circular orbit at 225-km altitude. Sampled from NRLMSISE-00 for a location above the equator at 12:00 UTC December 15, 2011. }
\begin{center}
\begin{tabular}{ll}
\hline\hline
Parameter & Value\\
\hline
$T_\infty$ & 809.2 K\\
$T_W$ & 300 K\\
$V$& 7770 m/s\\
$\rho_\infty$&$9.06\times10^{-14}$ kg/m$^3$\\
$q_\infty$&2.735$\times10^{-9}$ N/m$^2$\\
\hline\hline
\end{tabular}
\end{center}
\label{tab:PGcond}
\end{table}

\begin{figure}[htbp]
\begin{center}
\includegraphics[width=5in]{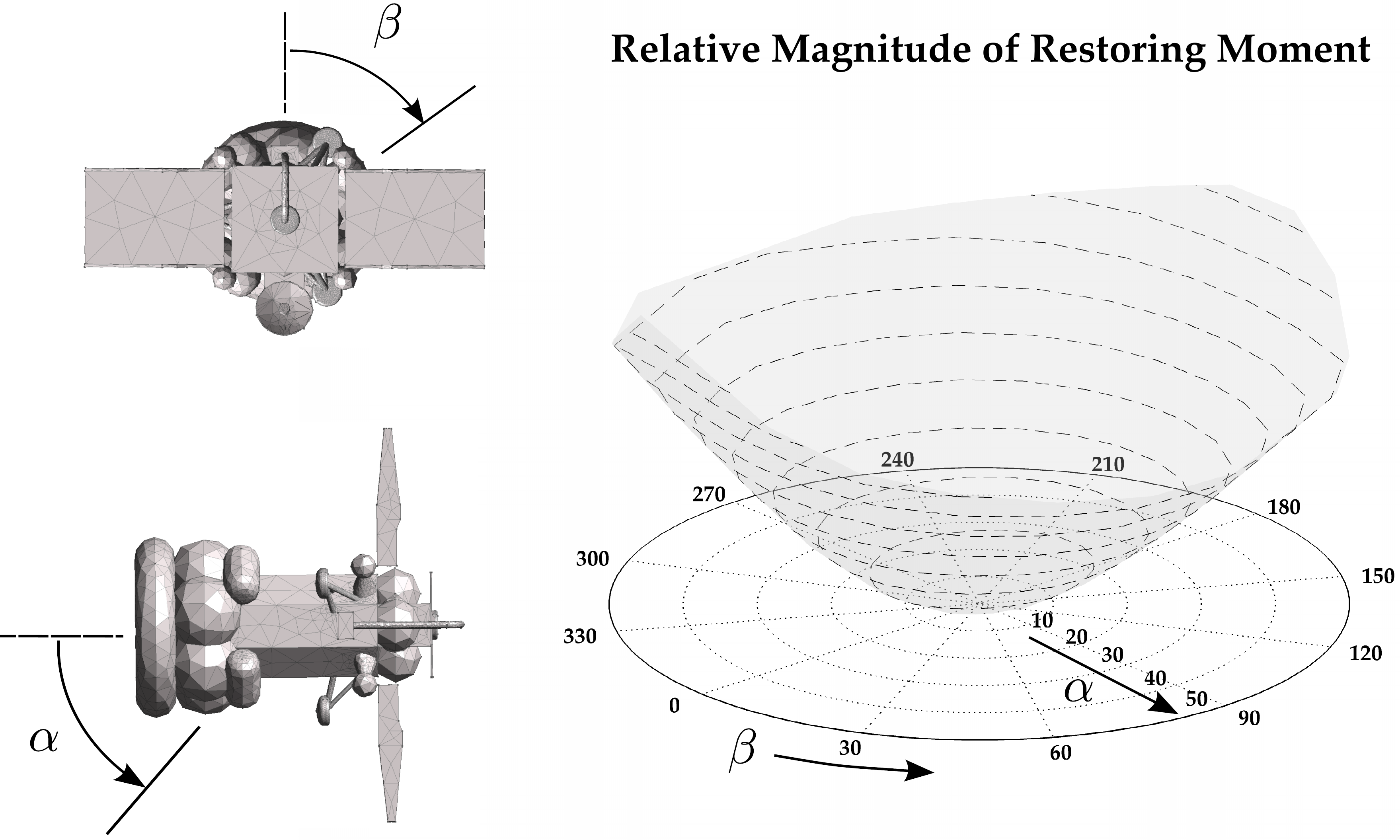}
\caption{The restoring moment---aerodynamic moment about $c/4$ ($c=5.5$ m) with positive direction decreasing the magnitude of $\alpha$---for the spacecraft model.  The moment is not axisymmetric about the roll axis, but it is monotonically increasing with increasing $\alpha$.}
\label{fig:restoring}
\end{center}
\end{figure}

To understand the stability of the spacecraft, we vary the angle of incidence about two axes and calculate the moment about the axis that would induce pitch; see Figure \ref{fig:restoring}.  This calculation reveals that for increasing angle $\alpha$ there is an increasing moment that serves to decrease the magnitude of $\alpha$.  As shown in the figure, this characteristic moment is roughly symmetric and produces a stability point very close to $\alpha=0^\circ$.  There is however asymmetry as we rotate about the roll angle $\beta$.

\begin{figure}[htbp]
\begin{center}
\includegraphics[width=2.8in]{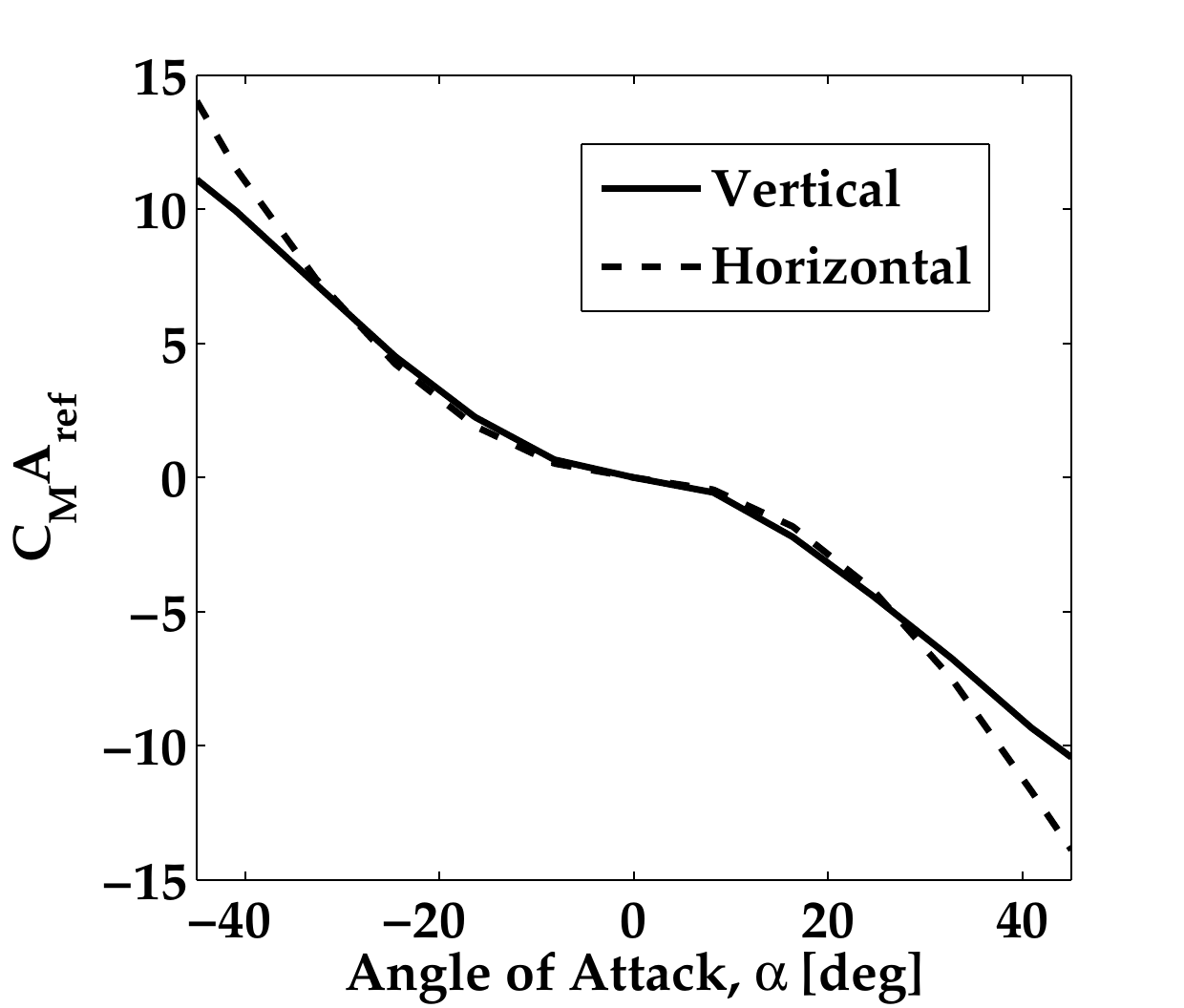}
\caption{Calculation of a pitching moment coefficient about the quarter-chord ($c=5.5$ m) for two roll angles where the solar panels are parallel to the pitch axis (Horizontal) and perpendicular (Vertical). The center of mass may be near this point due to the presence of unburned fuel in the tanks at the one end of the craft.}
\label{fig:Cm_vs_alpha}
\end{center}
\end{figure}

To better understand the asymmetry we look at the spacecraft geometry.  Choosing two $\beta$-angles so that the line connecting the center of each solar panel is either ``Horizontal'' (parallel to the pitch axis) or ``Vertical'' (perpendicular to the pitch axis), we recast the moments in terms of a pitching moment coefficient. To do this, we divide the moment by the dynamic pressure $q_\infty$ and the chord length $c$, leaving the coefficient scaled by an unknown reference area.  Defining positive moments as anything that pitches the spacecraft up, we obtain the results in Figure \ref{fig:Cm_vs_alpha}.  The asymmetry of the moments appears at high angle of attack, and manifests itself as a decreased moment for the ``Vertical'' configuration.  We hypothesize that this decrease in moment coefficient is due to a shadowing of a solar panel by the tank end of the spacecraft as it increases angle of attack.  If the panel was not shadowed, the dominant role of the drag force (lacking a substantial lift force) in creating the free-molecular moment would mean the moment coefficient would be more axisymmetric about the roll axis.  This phenomenon therefore can only appear when complex bodies are simulated with the appropriate considerations for flow shadowing.

\begin{wrapfigure}[18]{R}{0.4\linewidth}
\includegraphics[width=2.5in]{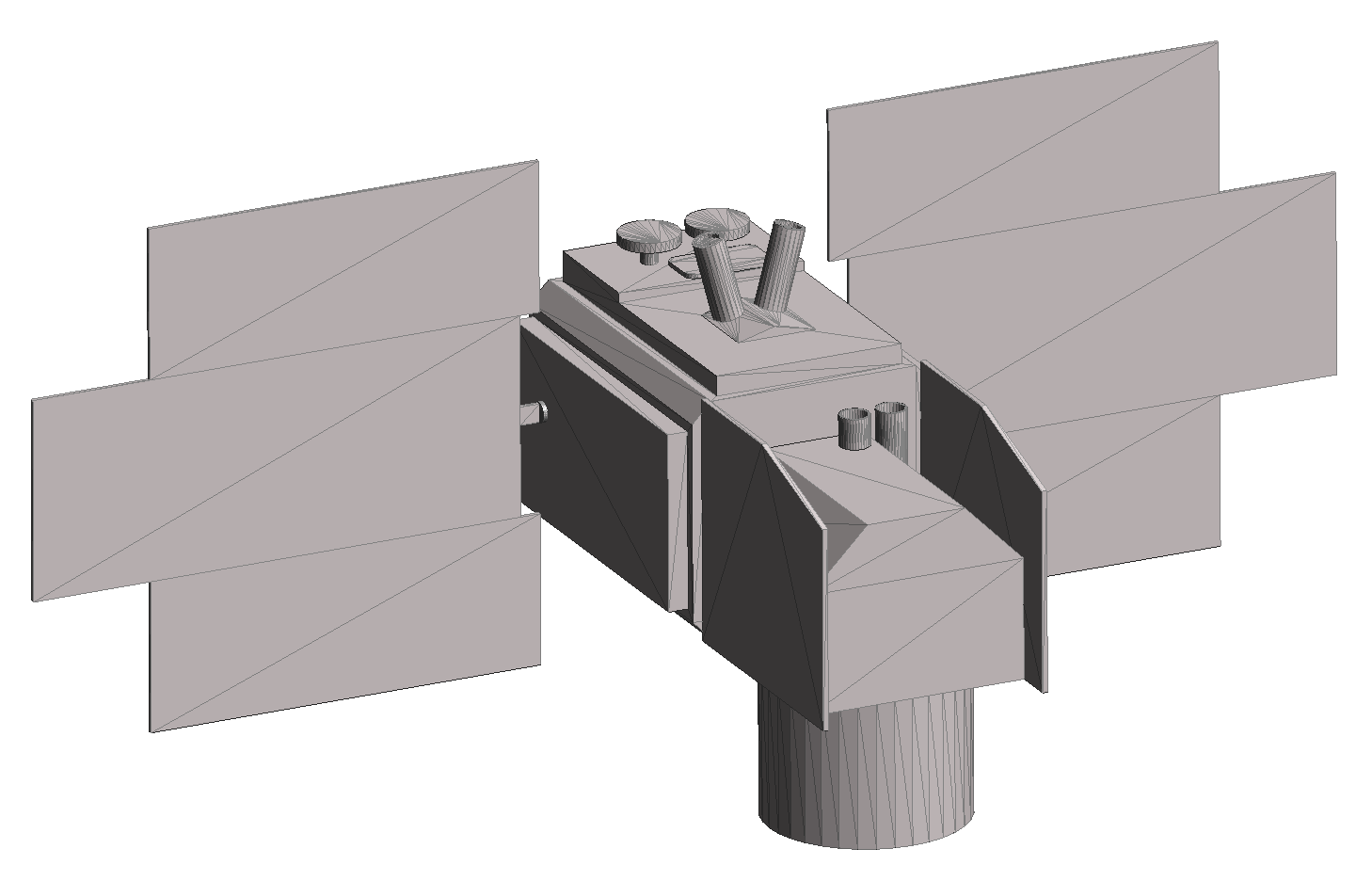}
\caption{Surface discretization used for ICESat simulations.  Solar panels are articulated.  For the Airplane mode calculations, we use the average of calculations with the panels perpendicular and parallel to the body axis.}
\label{fig:ICEsat_mesh}
\end{wrapfigure}

\subsubsection{ICESat pose estimation}

The ICESat science mission operated in two attitude modes during its mission: ``airplane'' and ``sailboat.'' These two modes, chosen to best present the solar panels to the sun, can clearly appear in derived force coefficients from the International Laser Ranging Service (ILRS) tracking data.  To show how this discrete attitude change is filtered through the drag-coefficient estimation with a typical orbit-determination routine, we create a model of ICESat (Figure \ref{fig:ICEsat_mesh}) and imposed the attitude constraints based on Webb's work \cite{webb2007}.  

\begin{figure}[htbp]
\begin{center}
\includegraphics[width=4in]{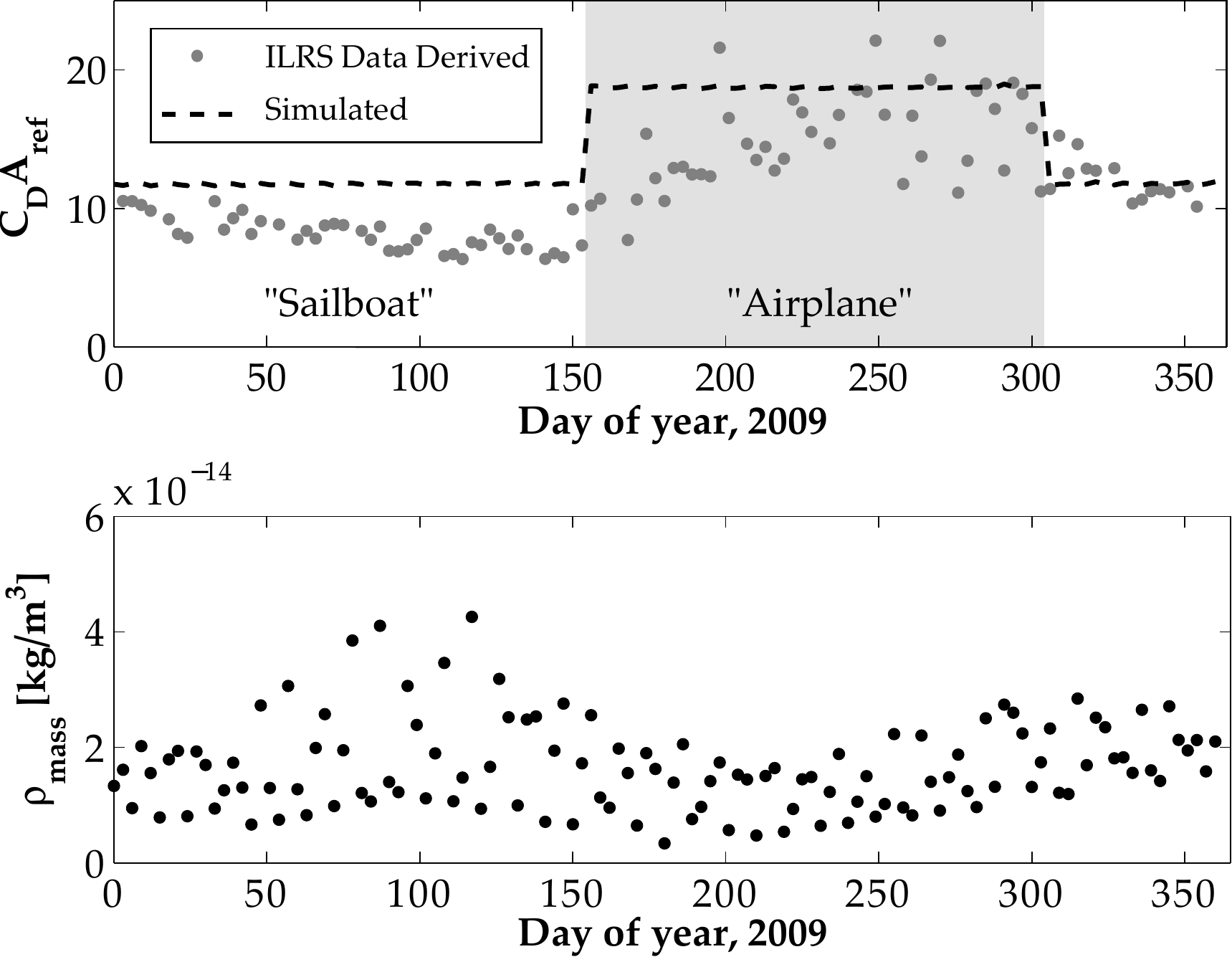}
\caption{ICESat $C_D$ as a function of the time in 2009. The drag coefficient is simulated using the on orbit velocities and atmospheric properties sampled from the NRLMSISE-00 model, and the second panel shows the atmospheric density from that model.}
\label{fig:ICEsat_paper}
\end{center}
\end{figure}

The orbit, along with the drag area $C_DA_{ref}$, is estimated with a least-squares filter that uses a three-day sliding window of laser ranges for its calculation.  The atmospheric conditions sampled from NRLMSISE-00 \cite{msis2000} and orbital velocity at the middle of each span are fed into the TPMC method to model a free-molecular drag coefficient with the TPMC method.  In Figure \ref{fig:ICEsat_paper}, we compare the estimated drag area to the modeled drag area.  We notice that the free-molecular drag area does not appreciably vary over the course of the year,despite changes in temperature of a few hundred Kelvin and density fluctuations by an order of magnitude.  The tracking data estimates, however, have a slow secular drift during ``sailboat'' mode and a wide scatter during the ``airplane'' portion of flight.  The simulation does allow us to precisely understand the step response of drag area filtered through tracking data, but the ILRS data only provides enough precision to roughly understand the physics of the aerodynamic forces on such a complicated body.

\section{Conclusion}
By demonstrating an improved approach to free-molecular aerodynamics, we hope to propel a stagnant field forward. The new physically consistent boundary condition ensures confidence in the calculation of forces and moments for a satellite model and with the power of modern workstation hardware, calculations become usable in a wider range of orbital analysis.  Beyond that, we provide verification tests (even accounting for concave bodies) for the TPMC method---often overlooked previously---and explore two intriguing applications in satellite dynamics. 

With this tool, we show that the uncontrolled dynamics of a complicated surface in the free-molecular regime may be dependent on accurate modeling of self-shadowing geometry, and prove aerodynamic stability for a spacecraft based on the Phobos-Grunt probe.  We can also extend the analysis drag forces with tracking data to complicated bodies with known geometry and attitude, such as ICESat. Although attitude and geometry are known for this spacecraft, the data provided by the ILRS service---some of the most precise tracking data available---when mixed with the NRLMSISE-00 ICESat atmospheric model is only accurate enough to qualitatively compare drag-force estimates to the physical model. This is enough, however, to understand large attitude changes with orbit determination methods based on metric tracking data. 

To summarize, we have shown the following:

\begin{itemize}
\item The TPMC method with its application to rarefied gas dynamics is ideal for implementation on new massively parallel high-bandwidth coprocessors such as GPUs.
\item We can derive set of boundary conditions that describe the correct distribution of states for influx particles that enter a spherical control volume.
\item TPMC codes can be verified appropriately using available analytical solutions.
\item TPMC codes can be used to calculate aerodynamic stability of satellites in the free-molecular regime.
\item We obtain qualitative agreement between free-molecular aerodynamic methods and real-world satellite dynamics with the ICESat mission.
\end{itemize}

The analysis and tool resulting from this work promises that future work may include  physically correct dynamical simulation for attitude and orbital-state propagation. As a result, finding the force for satellite drag becomes a little less uncertain.

\section*{Acknowledgments}

This work is sponsored by the U.S. Department of the Air Force under Air Force Contract FA8721-05-C-0002. Opinions, interpretations, conclusions and recommendations are those of the authors and are not necessarily endorsed by the United States Government. The first author wishes to express special thanks to David Gonzales for many years of support and insightful discussions about satellite dynamics and rarefied gases.  He also extends his thanks to MIT Lincoln Laboratory for supporting him through the Lincoln Scholars Program.

\bibliographystyle{aiaa-doi}	
\bibliography{./Library}  

\end{document}